\documentclass[twocolumn, 10pt]{IEEEtran}

\usepackage{hyperref} % create hyperlinks
\usepackage{graphicx}
\usepackage{amssymb}
\usepackage{amsmath}
\usepackage{mathtools}
\usepackage{cite}
\usepackage{stfloats}
\usepackage{subfigure}
\usepackage[font=small,labelfont=bf]{caption}
\usepackage{psfrag}
\usepackage[mathscr]{euscript}
\usepackage{acronym}  % make an acronym
\usepackage{booktabs} 
\usepackage[table]{xcolor}
\usepackage{multirow}
\usepackage{rotating}
\usepackage[skip=0.333\baselineskip]{caption}

\begin{document}
	
	\newcommand{\paperTitle}{Title}
	
	%---------------------------------------------------------------------------%
	%                     title, title footnote, header                         %
	%---------------------------------------------------------------------------%
	
	%\twocolumn
	
	% paper title
	\title{Is Blockchain Suitable for Data Freshness?\\
		- Age-of-Information Perspective}
	% \title{Distributed Secrecy in \\Multilevel Wireless Networks}
	
	% author names, IEEE memberships, corresponding address, title footnote %
	%\author{
	%%%%%%%% [begin] %%%%%%%%
	%	\vspace{0.2cm}
	%        Sungho~Lee, Minsu~Kim, Jemin~Lee and
	%        Ruei-Hau Hsu
	%[-0.5em]
	%
	%
	%
	%}
	\author{
		%%%%%%%% [begin] %%%%%%%%
		%	\vspace{0.2cm}
		Sungho Lee, Minsu Kim, Jemin Lee, \textit{Member, IEEE},%and 
		\\
		Ruei-Hau Hsu, \textit{Member, IEEE},  and Tony Q. S. Quek, \textit{Fellow, IEEE} 
		
		\thanks{
			Corresponding author is J. Lee.
			
			S.\ Lee, M.\ Kim, and J. Lee are with    
			Daegu Gyeongbuk Institute of Science and Technology (DGIST),  
			333, Techno Jungang-daero, Daegu, Republic of Korea 42988
			(e-mail: \texttt{\{seuho2003, kms0603, jmnlee\}@dgist.ac.kr}).
			
			R.\ H.\ Hsu is with National Sun Yat-sen University, 70 Lienhai Rd., Kaohsiung 80424, Taiwan, R.O.C.
			(e-mail: \texttt{rhhsu@mail.cse.nsysu.edu.tw}). 
			
			T.\ Q.\ S.\ Quek is with Information Systems Technology and Design Pillar, 
			Singapore University of Technology and Design, Singapore 487372 
			(e-mail: \texttt{tonyquek@sutd.edu.sg}).
		}
	}
	
	%% make the title area
	%% Don't write page number 0 to the cover page.
	\maketitle %% make the title area
	%% Don't write page number 0 to the cover page.
	
	%
	% \markboth{Submitted to IEEE Journal on Selected Areas in Communications}{\title}
	
	%
	%%%%%%%%% uncomment this section for a 2-column formt %%%%%%%
	%%%%%%%%% [begin] %%%%%%%%
	%\thispagestyle{empty}
	%  \textcolor{blue}{\framebox{\textsf{\small{Today: \today}}}}\\
	
	%
	%\newpage
	%%%%%%%%% [end] %%%%%%%%
	\setcounter{page}{1}
	\acresetall
	%%---------------------------------------------------------------------------%
	%%                           abstract and key words                          %
	%%---------------------------------------------------------------------------%
	\begin{abstract}
		Recent advances in blockchain have led to a significant interest in developing blockchain-based applications. While data can be retained in blockchains, the stored values can be deleted or updated. 
		From a user viewpoint that searches for the data, it is unclear whether the discovered data from the blockchain storage is relevant for real-time decision-making process for blockchain-based application.
		The data freshness issue serves as a critical factor especially in dynamic networks handling real-time information.
		In general, transactions to renew the data require additional processing time inside the blockchain network, which is called ledger-commitment latency. Due to this problem, some users may receive outdated data. As a result, it is important to investigate if blockchain is suitable for providing real-time data services. 
		In this article, we first describe blockchain-enabled (BCE) networks with Hyperledger Fabric (HLF). Then, we define age of information (AoI) of BCE networks and investigate the influential factors in this AoI. Analysis and experiments are conducted to support our proposed framework. Lastly, we conclude by discussing some future challenges.
	\end{abstract}

	\section{Introduction}
	Ever since Satoshi paved the way for blockchain technology, Bitcoin is considered to be the most popular blockchain-based application, and many researchers have actively studied the underlying architecture sustaining cryptocurrency. One of the outcomes is to utilize the blockchain concept for other non-monetary applications to ameliorate distributed systems and security issues of some emerging services.
	For example, an integrated blockchain platform for IoT devices, which are deployed worldwide, is proposed in [1]. The purpose of this platform is to provide the device owners with a practical application that offers comprehensive and immutable log, and allows easy access to their devices without the security concerns that stem from centralized IoT platforms. As another example, a blockchain system for 5G ultra-dense networks is established to redeem the shortcomings of the existing 4G authentication [2]. The authentication results are shared among a trusted access point group through the blockchain message propagation mechanism [2].
	%For example, a novel integrated blockchain platform for IoT devices, which are deployed worldwide, is proposed in order to provide the device owners with a practical application that offers a comprehensive, immutable log and allows easy access to their devices deployed in different domains with the characteristics of general IoT systems preserved, adducing the concerns of existing IoT platforms such as a cyber-attack and single point of failure that stem from their highly centralized architectures \cite{LeiKim:19}. As another example, a blockchain system for 5G ultra-dense networks is designated to redeem the shortcomings of existing 4G authentication and key agreement algorithm that fast and frequent authentication is difficult to be adapted \cite{CheCheXuHu:18}. The authentication results are shared among a trusted access points group through the blockchain message propagation mechanism \cite{CheCheXuHu:18}.
	
	Unlike stationary networks, there are dynamic networks that are required to handle real-time data. In such cases, \emph{data freshness} which is an indication of how new or relevant the current retained data is. During decision-making process, a bad decision may be made based on outdated retained data in a dynamic network.
	For instance, temperature data recorded by heat sensors has to be fed to the central controller in a timely fashion to sense whether there is an outbreak of fire for fire detection. Another example is a vehicular network, where vehicle location information needs to be regularly updated to the control center in order to make useful traffic management.
	
	How do we then estimate the degree of data freshness? Unfortunately, with the traditional performance metrics like delay and throughput, it is difficult to define. To be specific, let us assume that a source generates and transmits data via a channel according to first come first served (FCFS) scheduling. Packets that have not yet delivered will be queued at the source when the channel is busy. In this case, we may be able to achieve higher throughput by transmitting data more frequently. Nevertheless, this may not work well in terms of data freshness because the newly generated data will be backlogged and outdated due to the constrained bandwidth of the system. In an analogous way, although delay can be indeed minimized by transmitting data less frequently, a lower sampling rate results in a lack of update data and makes the data stale [3]. Thus, the traditional performance metrics have limitations on estimating the degree of data freshness.
	To measure the data freshness degree, the concept of age of information (AoI) has been recently introduced [4][5]. 
	When the generated data at a source (e.g., a sensor) is transmitted to a receiver (e.g., a monitor) for updating the data,
	the AoI at the receiver is defined as the time elapsed since the generation of the last successfully received update data. Hence, larger AoI means less fresh (i.e., more outdated) information.

	Although blockchain can provide more reliable management of data by guaranteeing data integrity, 
	there exists an additional delay to process data. 
	Therefore, it is questionable whether blockchain is suitable for dynamic networks,
	where sufficiently fresh data is expected. 
	In this article, to explore this unresolved problem, we exploit and analyze AoI for blockchain-enabled (BCE) networks. 
	Specifically, this article aims to 1) investigate influential factors on the AoI in BCE networks, 
	and 2) provide how to design BCE networks for reliable services of AoI-sensitive applications of dynamic networks.

	\section{Blockchain-enabled Networks}
	\subsection{Private Blockchain}
	A public blockchain, as its name indicates, allows any users to involve themselves in the network without requiring authorization despite of a demand for a strong security level. This rigorous requirement in terms of security confines the range of available consensus protocols for public blockchains to the proof-of-work (PoW) protocol. On the contrary, a \emph{private blockchain} refers to a blockchain platform for identified users, organizations, and entities only. One of the advantages of this platform is to enable equipping of power-efficient consensus protocols instead of resource-consuming consensus protocols due to the strict membership rules of private blockchains. Next, we will introduce Hyperledger Fabric (HLF), which is one of the most popular private blockchain platforms.
	
	\begin{figure*}[t!]
		\centering
		\captionsetup{justification=centering}
		\begin{center}   
			{
				\includegraphics[width=2.00\columnwidth]{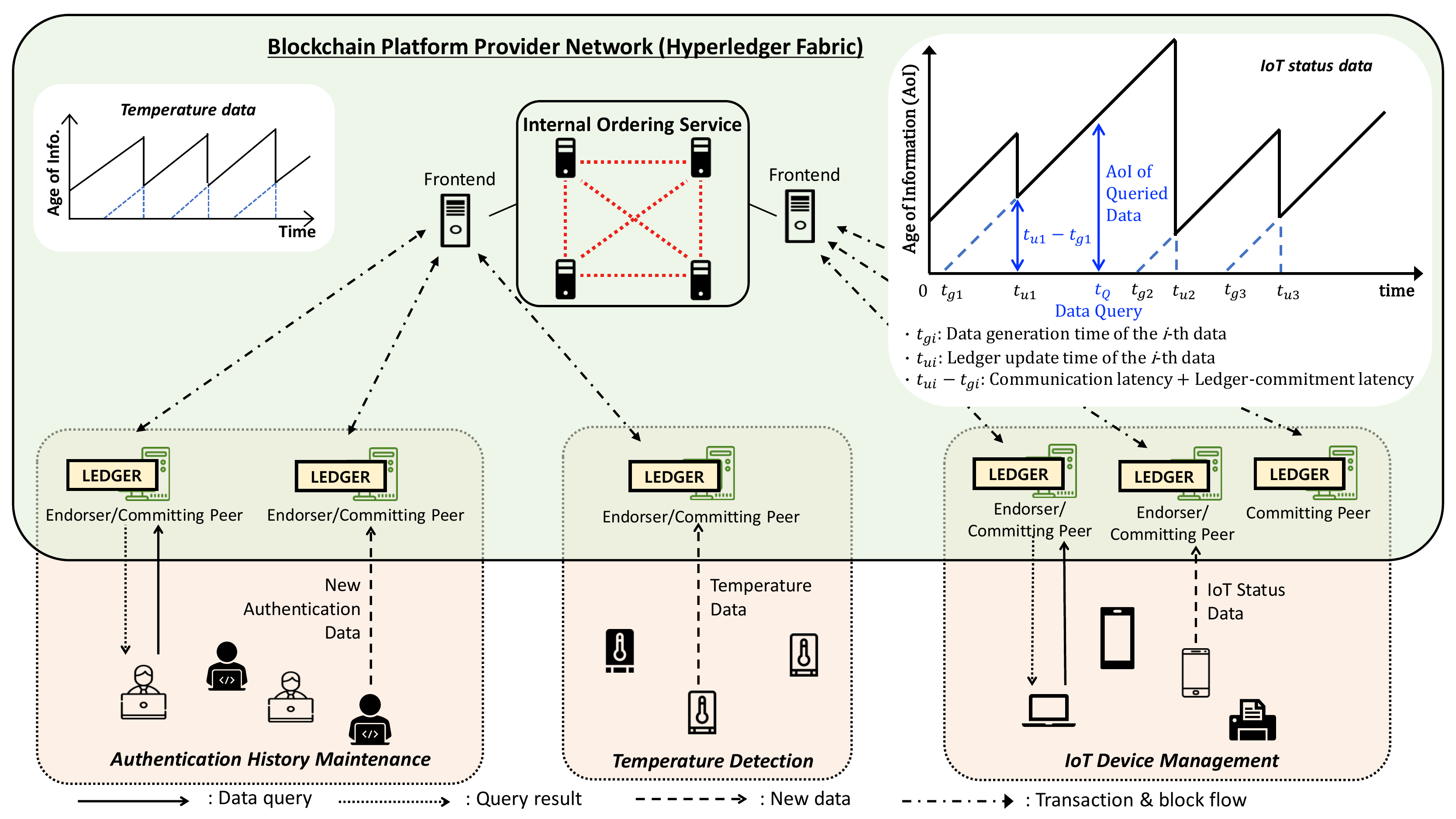}
				%			\vspace{-10mm}
			}
		\end{center}
		\caption{
			The structure of a blockchain-enabled network.
		}
		\label{fig:BCE_Structure}
	\end{figure*}

	\subsection{Hyperledger Fabric}
	Hyperledger Fabric (HLF) is an open source blockchain platform under the auspices of the Linux Foundation [6]. Fabric is not only to develop a private modular blockchain platform, but also to enable various smart contracts for wide use cases. Unlike general public blockchains, HLF has the unique structure to prohibit copied ledgers from diverging with supporting non-deterministic computer languages [6]. On the other hand, public systems are possible to cause a blockchain fork, prompting the usage of a deterministic programming language [6]. This distinct structure of HLF is closely related to the transaction flow, which is divided into three phases as described below.
	
	\subsubsection{Endorsing Phase}
	The endorsing phase is the first step that a transaction proposal enters at the beginning. The peers, which are entitled to endorse requests and involved in this phase, are referred to as endorsing peers (i.e., endorsers). 
	Their roles are described as follows:
	\begin{itemize}
		\item Endorsing peers execute the proposal using the requested smart contract against their local status so that new values and the current key versions are prepared in a read/write set. The peers then respond to the client with simulation results. 
		\item Endorsing peers make sure that each of the local states in all peers is identical while they perform transaction simulation, in order to prohibit ledger divergence. In case that any of the returned results are different from one another, the client discards the responses and refuses to proceed. If they are identical, on the other hand, the client collects them as a transaction, and transmits it to the ordering service.
	\end{itemize}
	
	% to (1) execute the proposal using the requested smart contract (i.e., chaincode) against their local status so that a new value and the current state for the specified key is prepared in a read/write set. The peers then respond with their signed simulation results. (2) make sure the local state in every peer is identical while they simulate it for prohibiting a ledger divergence. In case that any of the returned results are different, the client that requested for the ledger modification discards the responses and refuses to proceed. If they are identical, on the other hand, the client collects them in the form of a transaction, and transmits it to the ordering service for the next step.
	
	\subsubsection{Ordering Phase}
	The ordering phase refers to a step in which every endorsed transaction is ordered chronologically. The ordering task is conducted per channel independently in a node cluster. Note that HLF provides the channel concept, in which a channel-specific ledger is only shared across the peers belonging to the same channel for data isolation and confidentiality. Empowered to amass newly generated transactions and to create blocks per channel, the nodes inside the cluster are generally called ordering nodes (i.e., orderers). The new block is delivered to all the peers in the corresponding channel. Contrary to the other components, the ordering service does not appertain to any organization to independently fulfill the task.
	
	The ordering service is classified under three implementations according to its internal consensus method: Kafka/ZooKeeper, Byzantine fault tolerant state machine replication (BFT-SMaRt), and RAFT. The Kafka/ZooKeeper is the first fault-tolerant ordering service developed on the basis of Apache Kafka, which is a software platform for high-throughput message management to ensure crash fault tolerance (CFT). 
	The BFT-SMaRt is a Java-based Byzantine fault tolerant protocol for distributed environments, and there is only unofficial version of HLF with BFT-SMaRt, which was launched in [7]. The RAFT ordering service, which is available since 1.4.1v of HLF, is based on a leader-follower model (i.e., RAFT protocol).
	%The ordering service is classified under three implementations according to its internal consensus method: Kafka/ZooKeeper ordering service, BFT-SMaRt ordering service, and RAFT ordering service. \red{Note that whichever implementation is adopted, the intrinsic operation principles (i.e., block size and block-generation timeout parameters in Section IV) are preserved. The Kafka/ZooKeeper ordering service is the first fault-tolerant ordering service developed on the basis of Apache Kafka, which is a software platform for high-throughput message management to ensure crash fault tolerance (CFT) \cite{GorLeeGolKes:19}. This implementation method is the most general at present.} While it is vulnerable to Byzantine faults, an unofficial version of HLF with Byzantine fault tolerant state machine replication (BFT-SMaRt) was launched in \cite{SouBesVuk:18}. The RAFT ordering service, which is available since 1.4.1v of HLF, is based on a leader-follower model (i.e., RAFT protocol). \red{In this article, we focus on the prevalent ordering service type, that is, the Kafka/ZooKeeper implementation.}
	
	\subsubsection{Validation Phase}
	The block sent from the ordering cluster enters the last processing step, that is, the validation phase. 
	This step is mainly composed of two sequential verifications: (1) Validation system chaincode (VSCC) verification and (2) Multi-version concurrency control (MVCC) verification. 
	The \emph{VSCC verification} is to investigate whether the endorsement signature set in the transaction is valid or not. If the set does not satisfy the endorsement policy, the request is not only deemed to be invalid, but also banned from updating the ledger.
	The \emph{MVCC verification} is to compare the current versions of the keys captured 
	during the endorsing phase to those in the current states of the ledger, as stored locally by the peer [6]. All data in HLF are stored in the key-value scheme, in which each key of data works as a distinct identifier. A change in the version of a key arises in every data update. This principle implies that if both key versions are different, the data was already changed ahead of this transaction. Therefore, the transaction is not only marked as invalid, but also impossible to update the ledger in order to prohibit a blockchain fork [8]. If both versions are equal, on the contrary, the node then writes the new value to its local ledger and commit the block as the latest one.
	%The \emph{MVCC verification} is to compare the version of the key captured by the endorsing peers during the endorsing phase to those in the current state of the ledger, as stored locally by the peer \cite{AndBarBorCac:18}. A change in the version of a key arises in every data update. If the both version values are equal, the node then writes the new value to its local ledger and commit the block as the latest one. If they are different, on the contrary, the transaction is not only marked as invalid, but also impossible to update the ledger \cite{ThaNatVis:18}.}
	
	\section{AoI in Blockchain-enabled Networks}

	\subsection{Blockchain-enabled Network Structure}
	At this point, we state the definition of BCE networks. The BCE network is a particular network in which any service can be provided on the basis of blockchain technology. In other words, any application on the network can be implemented and underpinned based on blockchain in order to furnish users with a specific service.

	In Fig. 1, the blockchain platform provider not only facilitates service providers to exploit blockchain for their applications (i.e., authentication history maintenance, temperature detection, and IoT device management), but also helps them to manage and maintain data of their underlying system databases.
	%In other words, they can leverage blockchain technology for their service platforms conveniently, with the help of the blockchain platform provider so that their users can be assured of taking advantage of blockchain, such as data integrity and auditability.
	A new transaction proposal to update the ledger from an external device is transmitted to each assigned endorser to be granted permission.
	%\red{The endorser simulates the transaction proposal against the ledger upon verifying the user identity. The simulation result is conveyed to the ordering service after the client who sent the proposal checks it out. Note that every simulation result must be identical if multiple endorsers are assigned.} The ordering service, where new requests are assembled, generates new blocks per each HLF channel for privacy protection. The internal nodes of the ordering service collect them through the frontends. The delivered new blocks are allowed to update each ledger if they are successfully validated.
	For example, in the IoT device management application network in Fig. 1, an IoT device transmits status data in the form of a transaction proposal to the assigned endorsers. The endorsers simulate the proposal against their own ledgers, then the simulation results are delivered to the IoT device. If the simulation results are identical to each other, the proposal is conveyed to the ordering service in the form of a transaction. The ordering service, where new transactions are assembled, generates a new block per each HLF channel for privacy protection. The new block, which is delivered to the peers, is allowed to update each ledger with the new IoT status data included in the block if the transaction is successfully validated.

	%\begin{figure}[t!]
	%	\centering
	%	\captionsetup{justification=centering}
	%	\begin{center}   
	%		{
	%			%
	%			\includegraphics[width=1.00\columnwidth]{figures/ageGraph.pdf}
	%			%			\vspace{-10mm}
	%		}
	%	\end{center}
	%	\caption{
	%		Sample path of the AoI in a BCE network.
	%	}
	%	\label{fig:Age}
	%\end{figure}
	
	\subsection{AoI Elements in Blockchain-enabled Networks}
	Expanding the AoI concept to the BCE network requires a different context from traditional networks. The AoI in a traditional network is defined as the elapsed time since the packet for the last update was generated. This definition mainly considers communication latency and queueing-based processing latency. However, in the BCE network, an additional delay, which is not queueing-based, needs to be taken into account because the series of data processing in blockchain also delay updating data. Therefore, we need to consider the time spent in the blockchain (i.e., ledger-commitment latency) as well as the communication latency. One of the main challenges is that real-time applications (e.g., sensor networks, spectrum sharing networks, and vehicle location management network) [1][9] may not be able to provide the latest information for their users due to fast dynamic data. Out-of-date information may incur inaccurate outputs and unintended operations. This problem can be shown clearly in Fig. 1.
	%
	%\red{Those applications may not be able to provide 
	%the latest information to users due to the data dynamics, which may incur undesired decisions, inconsistent outputs, and unintended operations. 
	%This problem can be shown clearly in Fig. 2.}
	%The applications with those kinds of data may not be able to provide their users with the latest utilizable information as query results due to their dynamic properties, which may incur undesired decisions, inconsistent outputs, and unintended operations. 
	
	In Fig. 1, a sample path of the AoI for IoT status data is illustrated.
	%The sample path of the AoI in those BCE networks is described in Fig. 2. 
	Since the last update, the AoI is increasing linearly. When a new update occurs at the source at time $t_{gi}$, the new data, which will be committed to the blockchain, takes some time to reach the target blockchain network and becomes effective through the processing steps indicated in Section II-B. The decrease in the AoI at a right angle coincides with the ledger-commitment at time $t_{ui}$. Note that the AoI does not touch zero because the data has the update latency (e.g., $t_{ui}$-$t_{gi}$) as its AoI until committed to the ledger. For clarification on the proposed topic, we instantiate the elements of AoI in the BCE network in this section.

	\subsubsection{Data Generation Frequency and Time Distribution} %Time and Frequency}
	The data generation time is closely related to the data update time at the ledger. In the sample AoI path in Fig. 1, when a packet is generated at time $t_{g1}$ and the block containing the packet is completely committed and updated at time $t_{u1}$, the AoI of the data in the ledger is reset to $t_{u1}$-$t_{g1}$. The AoI then starts to linearly increase again until the next data update at time $t_{u2}$. Hence, the AoI depends on two factors: the data generation frequency and the data generation time distribution. The generation frequency is related to the average number of new packets generated per second. The generation distribution can be modelled in a stochastic manner (e.g., exponential distribution) or a deterministic manner (e.g., periodic generation).
	%The data generation time is closely related to the data update time at the ledger. For instance, in Fig. 2, when a packet is generated from the source at time $t_{g1}$ and the block containing the packet is completely committed and updated at time $t_{u1}$, the age of the data in the ledger is reset to $t_{u1}$-$t_{g1}$. The age then starts to linearly increase again until the next data update at time $t_{u2}$. With respect to every moment at which a new packet is created (e.g., time $t_{gi}$), the age depends on two factors, namely, the data generation frequency and the data generation distribution. The generation frequency is related to the average number of new packets generated per second. The generation distribution can be stochastically modelled or even generated in a deterministic manner.
	
	\subsubsection{Communication Latency}
	The communication latency is the time taken to transmit data to the target blockchain network. The time-stamped packet, which is generated by the source, is conveyed to the associated base station or an access point. The data is then forwarded to the BCE network to update the ledger.
	
	\subsubsection{Ledger-commitment Latency}
	The ledger-commitment latency refers to the time taken to process a transaction in the BCE network. When the request containing fresh-data successfully arrives at the target blockchain, the data needs to go through the transaction-processing steps, indicated in Section II-B.
	
	From those elements, affecting the AoI, we can also analyze the AoI in various aspects. 
	Once the distribution of the ledger-commitment latency is analyzed (which can be the most challenging part), 
	the average AoI and the complementary cumulative distribution function (CCDF) of AoI can be obtained based on the results in [4] and [5].
	%The ledger-commitment latency refers to the time taken to process a transaction in the BCE network. As the request containing fresh-data from a data source successfully arrives at the target blockchain network, the data is not committed to the ledger promptly, but goes through the transaction-processing steps as indicated in Section II-B.
	
	%This delay may vary with many factors. The consensus protocol type, for example, can affect this latency since the protocols have different operations from one another. A PoW system is inevitably employed in solving the hash puzzle for mining, while a BFT-SMaRt system requires enough time to communicate among the internal peers for consensus on a new block. The other influential factors are also introduced in detail in Section IV-A.
	
	%This delay may vary with the consensus protocol type selected for the system because of the different operations among the protocols. For instance, generally speaking, a blockchain with the Proof-of-Work (PoW) consensus protocol will spend some time to solve the hash puzzle in an attempt to mine a new block into which the transaction is pushed. Besides, a Practical Byzantine Fault Tolerance (PBFT) based system will also require enough time to communicate among the internal peers for consensus on a new block. This latency definitely implies that no state update happens until the involved peers finish the consensus step and block-commitment according to the accepted procedure. In this regard, it is of importance for information freshness to give a thought to this processing time.
	
	\begin{figure}[t!]
	\centering
	%\captionsetup{justification=centering}
	\begin{center}   
		{
			\includegraphics[width=1.00\columnwidth]{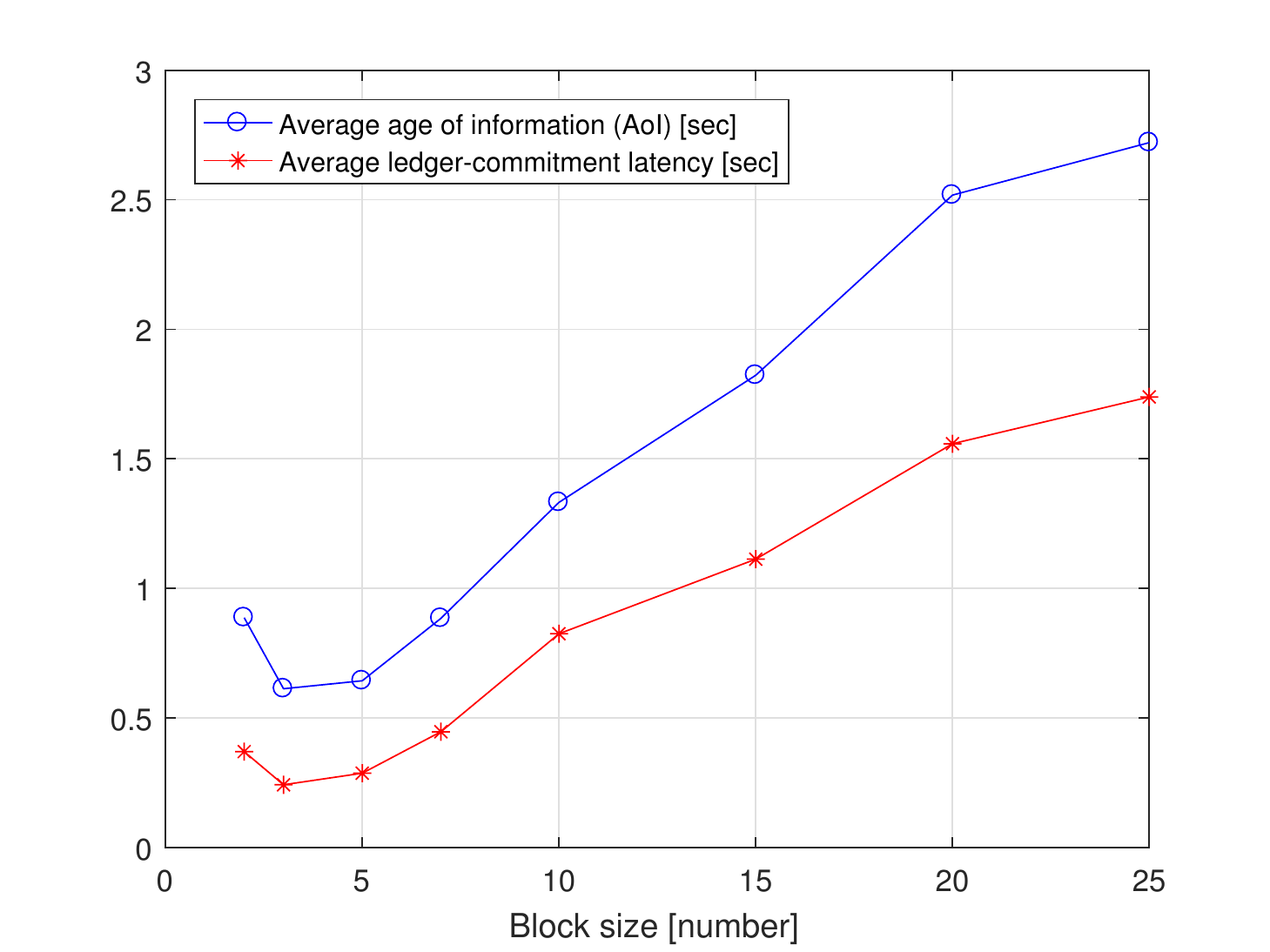}
			%			\vspace{-10mm}
		}
	\end{center}
	\caption{
		Effect of the block size on the average AoI and latency, where the block-generation timeout is 2 seconds, and the ratio of the target key transactions is 30\%.
	}
	\label{fig:Blocksize}
	\end{figure}

	\section{Influential Factor Analysis}
	%With consideration for the importance of the AoI in the BCE network, we ponder three influential factors to retain data freshness maximally from various aspects and analyze their impacts in this section. 
	%To this end, we also
	In this section, we provide experiment results by implementing HLF and show influential factors on the AoI. 
	For the implementation of ordering service, we use Kafka/Zookeeper since it has been most widely used and considered to be more stable than other implementation methods, such as BFT-SMaRt and RAFT.
	We construct a blockchain platform provider network with all necessary HLF components using HLF 1.3v [10]. Note that the insights obtained in this section can also be applied to HLFs with the other implementation methods of the ordering service since different ordering service only changes how to arrange transactions in a block. The Kafka/ZooKeeper ordering service cluster consists of 4 Kafka nodes with one frontend node. Besides, the blockchain network has one HLF channel, consisting of one endorsing peer and two committing peers, which are only allowed to commit blocks.
	%In the following, we provide experiment results by deploying a Kafka/ZooKeeper ordering service cluster \red{consisting four Kafka nodes with one frontend node. For this experiment, we construct a blockchain platform provider network with all necessary HLF components using HLF 1.3v. Besides, the blockchain network has one HLF channel consisting one endorsing peer and two committing peers which are only allowed to commit blocks.} %The physical machine on which the containers run is equipped with Intel\textsuperscript{\textregistered} Core\textsuperscript{TM} i7-6700 @ 3.40GHz and 2GB of memory.}
	The request (transaction) arrives at HLF regularly at the rate of 10 transactions/second.
	A certain percent of the arrived requests tries to update the target key-value. We focus on the AoI of this target data, and this configuration is for all experiments unless specified otherwise.
	
	%To explain their influence, we deploy the same Kafka/ZooKeeper ordering service cluster with one frontend node in HLF 1.3v for every experiment. The request arrival rate to the HLF network is fixed as 10/sec, while the inter-arrivals of them are distributed in a deterministic fashion. A certain percent of the generated requests try to update the target key-value. Only this target key is observed to depict the simulational results. On the other hand, the rest of them try to update other random keys. This detail configuration environment is for all experiments unless specified otherwise.
	
	\subsection{Blockchain Parameters}
	The blockchain parameters refer to blockchain network configuration. %The compelling fact is that their variations bring about remarkable changes in the performance. In HLF, for example, a client must await the responses from some appointed endorsing peers. This signifies that the number of endorsing peers, from which the client needs to receive endorsements, may have an impact on the performance. Besides, the ordering nodes comply with the pre-defined block size and block-generation timeout value while producing new blocks, which also may imply they have an influence upon the performance. We analyze the impacts of the two main parameters aforementioned, that is, block size and timeout, on data freshness.
	In HLF, the block size and the block-generation timeout significantly affect the performance, which we will investigate in the following.
	
	\subsubsection{Block Size}
	The block size parameter is the maximum number of transactions in a block. 
	When a transaction arrives in the ordering phase, it is included in a block and awaits others 
	until the number of transactions in the block becomes the block size or the block-generation timeout expires.
	Consequently, an ordering service with a larger block size results in longer waiting time (i.e., longer ordering latency),
	as the larger block requires more time to be full.
	%
	%When transactions are delivered to the network, they need to await the others to be included into a block. As a result, an ordering service with a larger block size will spend more time than one with a smaller block size. This is due to the fact that a larger block requires more time to be full.
	%
	%As a result, an ordering service with a larger block size setting eventually makes them spend more time than the one with a smaller block size setting. This is because the time taken to wait until the block becomes full always longer with a larger block size.
	%
	%The block size parameter is the maximum number of transactions in one block. As transactions are delivered to the network at a moderate rate, the transactions await for the others to be included into a block. In case of an ordering service with a larger block size setting eventually makes them spend more time than the one with a smaller block size setting. This is because the time taken to wait until the block beceomes full becomes always longer as the block size becomes larger. 
	%
	%
	Figure 2 demonstrates the impact of block size. We can see that the average AoI of the target key data increases continuously with the block size. However, when the block size is small (i.e., smaller than three), the AoI decreases with the block size. This unlooked-for result is obtained from the fact that the block generation rate at the ordering service is beyond the block-commitment rate at that point. As a result, new blocks stack up in each peer's waiting queue for commitment. Therefore, it is important to determine appropriate blocksize for shorter AoI in a BCE network.
	
	\begin{figure}[t!]
	\centering
	%\captionsetup{justification=centering}
	\begin{center}   
		{
			\includegraphics[width=1.00\columnwidth]{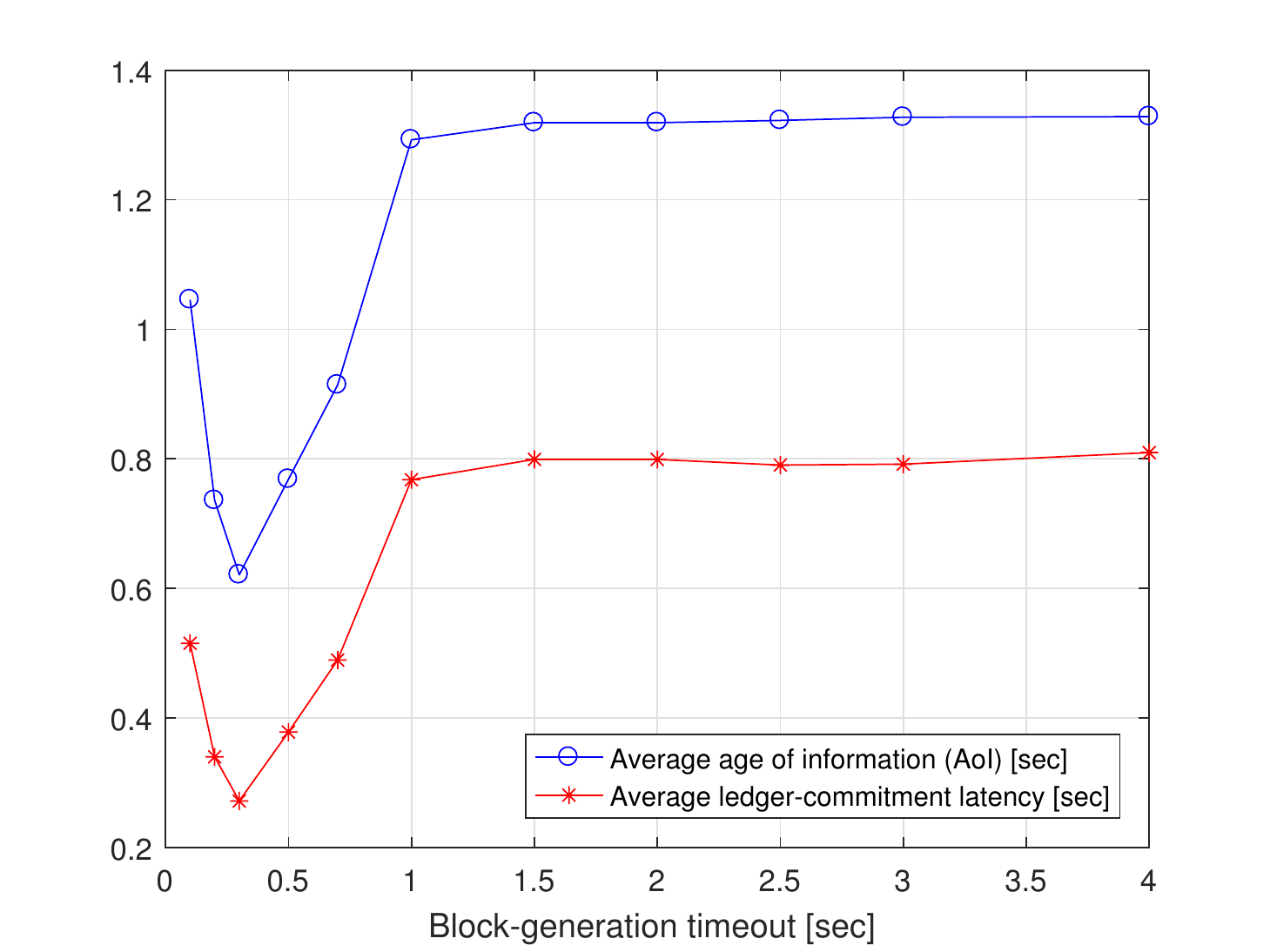}
			%			\vspace{-10mm}
		}
	\end{center}
	\caption{
		Effect of the block-generation timeout on the average AoI and latency, where the block size is 10, and the ratio of the target key transactions is 30\%.
	}
	\label{fig:Timeout}
	\end{figure}
	
	\subsubsection{Block-generation Timeout}
	The block-generation timeout refers to the maximum time that a transaction waits for the others in the ordering service. This parameter is used to avoid long latency by allowing a block to move to the next step even if the block has not been completely full. Hence, as the timeout increases, the blocks are generated slower (i.e., the ordering service latency increases), but the latency in validation phase generally decreases.
	Figure 3 illustrates the impact of block-generation timeout on the average AoI of the observed data. 
	When the timeout is short (e.g., less than 0.3 seconds), 
	we can see that the average AoI decreases with the timeout. 
	%This is because when the timeout is short, 
	This is because the decrease in the validation latency is more significantly than the increase in the ordering service latency for a short timeout range.
	On the other hand, when the timeout is greater than 0.3 seconds, 
	the increase in the ordering service latency becomes more significant, so the average AoI increases with the timeout.
	The average AoI becomes eventually saturated as the timeout becomes large (e.g., greater than 2 seconds) because transactions start to form a compacted block before the timeout expires, which makes the impact of the timeout disappear.
	
	\begin{figure}[t!]
	\centering
	%\captionsetup{justification=centering}
	\begin{center}   
		{
			\includegraphics[width=1.00\columnwidth]{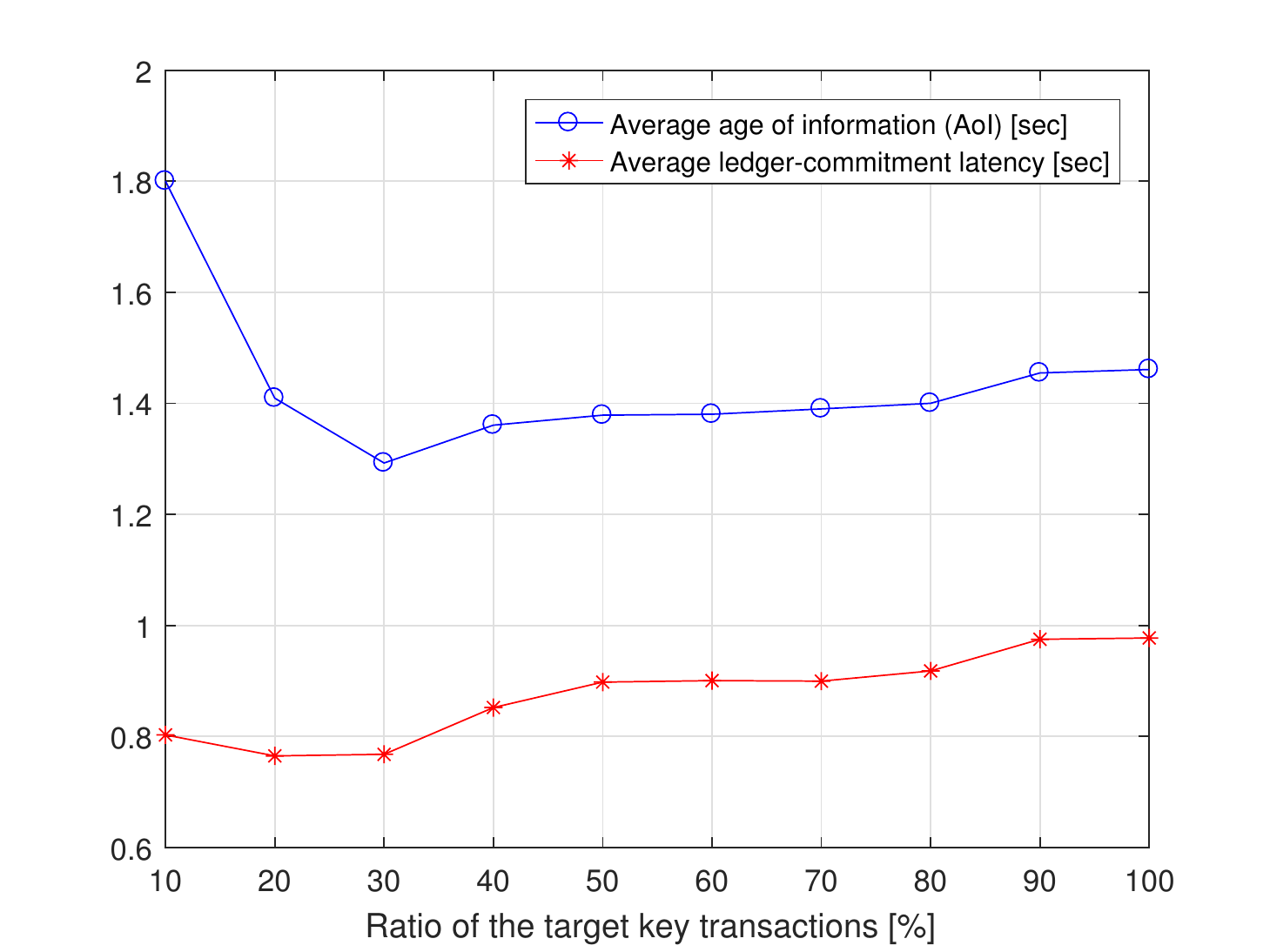}
			%			\vspace{-10mm}
		}
	\end{center}
	\caption{
		Effect of the data generation frequency on the average AoI and latency, where the block size is 10, and the block-generation timeout is 1 second.
	}
	\label{fig:Frequency}
	\end{figure}
	
	\subsection{Data Generation Frequency}
	When data is transmitted through a wireless channel, the other generated data will be on standby to be delivered and cumulate in the transmitter buffer. At this time, it is essential to tune the data generation frequency appropriately since excessively high or low generation frequency has negative influence on the AoI. As another consideration, furthermore, the lastcome first served (LCFS) discipline is worth considering for a packet transmission method than the FCFS discipline.
	This is because the information of lately generated packet is always fresher than those of packets waiting in the queue. 
	Hence, it is more preferable to transmit
	the newest packet first than the queued ones through the LCFS discipline. 
	It is also shown that the LCFS scheme can conserve
	data freshness better than FCFS
	discipline even at high arrival rate of packets in [11].
	
	Figure 4 shows the impact of ratio of the target key requests 
	(i.e., transactions attempting to update the target data) to the total requests.
	Note that the impact of communication latency is excluded here, and increasing the ratio is equal to increasing the generation frequency of the target key requests.
	When the ratio is small, 
	although the average ledger-commitment latency is relatively low due to less processing requests in HLF,
	the AoI is large due to a lack of information to update.
	As the ratio increases, the AoI also decreases, but the trend morphs into its opposite after 30 percent.
	%\red{The reason is from an increase in the possibility that the target key transactions arrive the ordering service relatively earlier than the others in each block. In one block, the early-arriving transaction needs to wait longer than the other transactions until the block is exported. 
	This is from an increase in the possibility that the captured version of the target key at the request generation is different from the current version in the ledger at the validation phase, 
	which is considered as invalid during the MVCC verification.  
	This is more likely to occur especially when the key requests generate frequently,
	since a new request captures the version with which the pre-generated one was already simulated,
	and the key version can be changed by the previous one before the new request is committed with it.
	%After the early-arriving one updates the data, moreover, the version of the key is changed. Consequently, the other transactions, which access the key with the same version subsequently, are impossible to update the data, and considered as invalid during the MVCC verification. 
	This result not only increases the average latency, but also hinders instantaneous data updates with fresher information.
	
	%, which the transactions considered as invalid for the same version contain.}
	%The reason of this situation results from an increase in the possibility that target key transactions arrive the ordering service earlier than the others relatively in each block. In one block, the early-arrived transactions are compelled to wait longer than the others until the block is exported. With an increase in the ratio of the target key transactions, they are delievered to the ordering serivce at higher rates gradually, exerting a bad influence on the latency. Moreover, during the validation phase, the subsequent target key transactions in the block cannot update the ledger in violation of the MVCC verification, even though they have fresher information than the precedent one. This trade-off relation definitely implies that the frequent generation of data does not always reduce the AoI in the BCE network.
	
	%possibility that first transaction in one block is committed by the FCFS discipline, which is a selection of the oldest transaction. The other transactions in the same block with fresher information which access the same key are eventually considered as invalid by the MVCC verification. Note that the mean latency does not decrease contrary to the shape of the average age. This fact underpins that it is not always better to thrust as many transactions as possible into the blockchain system.
	
	\begin{figure}[t!]
		\centering
		%\captionsetup{justification=centering}
		\begin{center}   
			{
				\includegraphics[width=1.00\columnwidth]{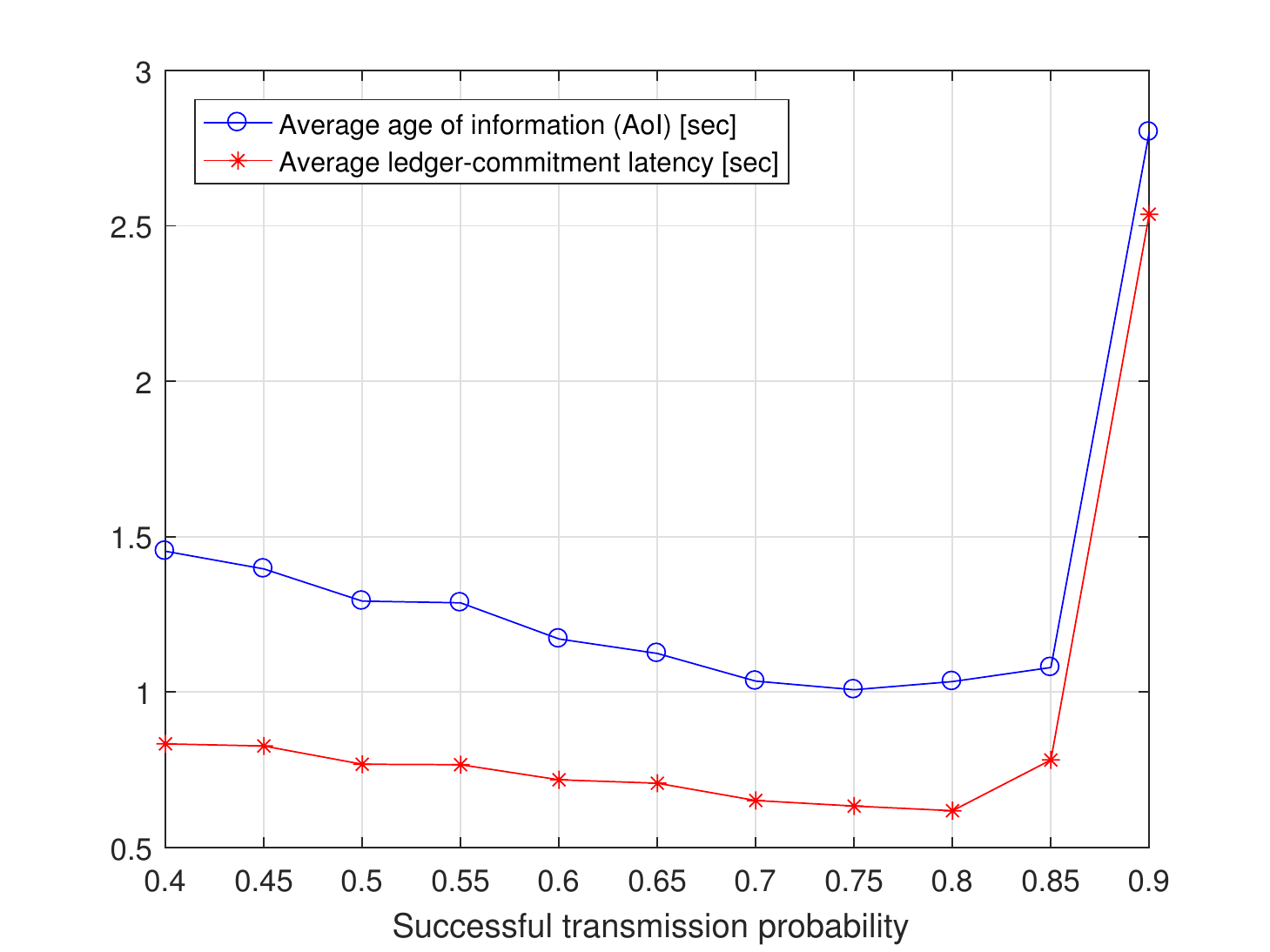}
				%			\vspace{-10mm}
			}
		\end{center}
		\caption{
			Effect of the successful transmission probability (STP) on the average AoI and latency, where the total data generation rate is 20 packets/second, the block size is 10, the block-generation timeout is 1 second, and the ratio of the target key transactions is 30\%.
		}
		\label{fig:STP}
	\end{figure}
	
	\subsection{Communication Parameters}
	The communication parameters refer to impactful elements on communication performance, 
	which also affect on the data freshness.
	
	\subsubsection{Scheduling Policy}
	The first element we focus on is the scheduling policy, which is referred to a rule to coordinate channel allocation. In a wireless communication system, where multiple nodes can transmit packets, it is necessary for base stations to allocate channels to each node properly. An unsuitable policy may not only incur interference protracting data updates, but also lead to unfair chances to transmit data to update. The biased channel allocation may increase the AoI of particular data in the ledger. Note that the conditions that scheduling policies have to satisfy can change depending on their network environments [12][13] such as the minimum throughput constraint of each node [12]. Therefore, a deliberate selection of scheduling policy must be given to minimize the AoI.
	
	%The communication parameters refer to impactful elements on communication performance. The first element we focus on is the scheduling policy, which is referred to a rule to coordinate channel allocation. In a wireless communication system, where multiple nodes can transmit packets, it is necessary for base stations to control the channel allocation to each node with a proper scheduling policy, in order to avoid interference that protracts data updates because an unsuitable policy is possible not only to incur higher AoI overall, but also to lead to unfair chances to transmit data on partial nodes. The nodes with rarely allocated channels have much higher AoI than that of the nodes with frequently allocated channels. Note that the conditions that have to be satisfied by the scheduling policy are not stationary, but depend on various network settings \cite{14}\cite{15}, for example, the minimum throughput constraint of each node is considered in \cite{14}. Therefore, a deliberate selection of scheduling policy must be given to minimize the AoI with consideration for the current network environment.
	
	\subsubsection{Transmission Power}
	The transmission power of a source is an important element, especially when battery-operated or energy-constrained sources are deployed [14].
	The successful transmission probability (STP), which is defined as the probability that the receiver receives the update data reliably, generally increases with the transmission power of the source. 
	As only successfully received data is used for the information update at the receiver, we expect to have lower AoI with higher STP. 
	This result is also shown in Fig. 5, which shows the average AoI with the STP.
	
	However, when the STP is greater than 0.8, the average AoI rapidly increases with the STP. 
	This unforeseen increase is from the long ledger-commitment latency. 
	Specifically, when $\lambda_{\text{t}}$ is the total data generation rate of sources and $\theta$ is the STP, 
	the transaction arrival rate of the HLF channel in the BCE network is defined as $\lambda_{\text{arr}}=\lambda_{\text{t}}\theta$. 
	As $\lambda_{\text{arr}}$ increases, blocks are generated faster at the ordering phase, while those blocks need to wait in a queue longer to be committed to the ledger since each block is validated serially in the validation phase. Hence, severely high transaction arrival rate can impede the block-commitment process, as also discussed in Section IV-A.

	\begin{figure}[t!]
	\centering
	%\captionsetup{justification=centering}
	\begin{center}   
		{
			\includegraphics[width=1.00\columnwidth]{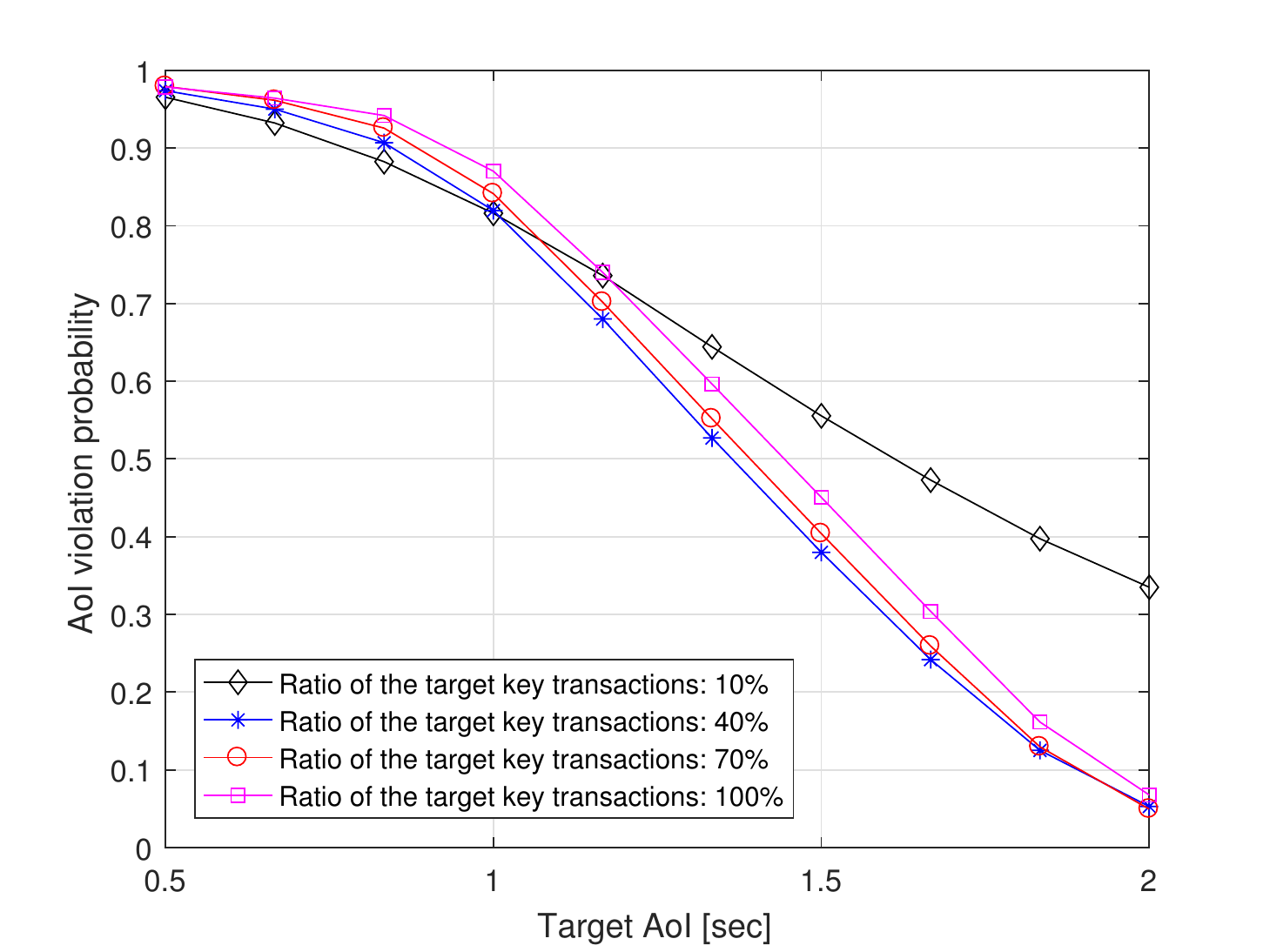}
			%			\vspace{-10mm}
		}
	\end{center}
	\caption{
		Effect of the target AoI on the age violation probability, where the block size is 10, and the block-generation timeout is 1 second.
	}
	\label{fig:violationProb}
	\end{figure}
	
	\section{Future Challenges}
	%With much attention in employing blockchain for extensive applications, the blockchain has been developing apace in various fields.
	As a prerequisite for successful blockchain utilization, the three influential factors are discussed with experiments in Section IV. We then now need to think about whether blockchain is suitable enough to provide fresh-data requiring services. Let us consider an application that has a target AoI for the system performance. In case of a sensor network, the degree of sensor accuracy decreases with an increase in the AoI of sensing data [15], and the target AoI can be given as the one that guarantees a certain level of sensor accuracy. Figure 6 shows the impact of target AoI on the AoI violation probability, which is the probability that the AoI is greater than the target AoI. When the target AoI is large (e.g., greater than 1.9 seconds), the AoI violation probability is less than 0.1, which is quite reliable. However, for small target AoI applications, BCE networks may not be suitable. In the rest of this section, we introduce network design aspects that can be considered for data freshness as future challenges as well as possible solutions to lower the AoI from each parameter's perspective.
	
	%To apply blockchain for a wide-range of applications, a fine-grained BCE network design is necessary to make the AoI violation probability smaller even for small target AoIs.
	
	\subsection{Additional Blockchain Parameters}
	Even though the impacts of main blockchain parameters are analyzed in Section IV, the other factors to consider still exist. We cover two additional blockchain considerations in this subsection.
	
	\subsubsection{Block Generation Rate}
	The block generation rate refers to the rate of generating a new block at the ordering service. 
	This parameter is not controllable directly, but largely affected by the others blockchain parameters.
	For instance, generally, the smaller block size, the shorter block-generation timeout, and the higher transaction arrival rate a BCE network has, the new blocks are generated faster. However, as also shown in Figs. 2 and 3, 
	the smallest block size and the shortest timeout do not give the lowest AoI.
	
	When the total arrived transactions during a certain period is $m$ and for $n_1<n_2$, the time taken to validate and commit $\frac{m}{n_1}$ blocks with size $n_1$ is not always less than the time taken to validate and commit $\frac{m}{n_2}$ blocks with size $n_2$, especially for a high transaction arrival rate and a small block size [8]. This is because new blocks flooding from the ordering service into peer buffers at a high rate can lead to longer latency due to the CPU-intensive operations. Therefore, it is important to properly adjust the block generation rate to avoid long latency.

	\subsubsection{Number of Nodes}
	The AoI can also be affected by the number of participating nodes in blockchain such as endorsing nodes and Kafka nodes of the ordering service in HLF.
		
	\begin{itemize}
		\item The endorsement policy is a guideline for peers to recognize a properly endorsed transaction [8]. This policy generally specifies by which endorsing peers transactions have to be endorsed. In this sense, users cannot help spending longer time to collect endorsements if multiple endorsing peers are compulsory. In our experiments, it is shown that the average AoI increases from 1.29 to 1.34 seconds as the number of endorsing peers increases from 1 to 3 when the Kafka nodes are four.
			
		\item The number of Kafka nodes, used in the ordering service, is connected with CFT, and more Kafka nodes can endure more node failures. For instance, it can allow up to one node failure for four Kafka nodes (i.e., the minimum number) and two node failures for five Kafka nodes. However, as the number of Kafka nodes increases, it requires more processing time. In our experiments, it is shown that the average AoI increases from 1.34 to 1.46 seconds as the number of Kafka nodes increases from 4 to 5 when the endorsing peers are three.
	\end{itemize}
	%\subsubsection{Endorsement Policy}
	%The endorsement policy is a guideline for peers to recognize a properly endorsed transaction using either \textsf{AND/OR} or \textsf{NOutOf} syntax \cite{ThaNatVis:18}. While validating a transaction in a block, the peers in the network judge its validity based on the endorsement policy defining by which endorsing peers transactions have to be endorsed. In this sense, users are required to spend longer time to collect endorsements if the policy is complicated. The endorsement set left by the endorsers serves as a proof that which endorsing peers did agree upon the transaction. Therefore, from the AoI viewpoint, it is important not only to make a choice between the both syntax forms, but also to decide the number of endorsing peers.
	
	%\subsubsection{Number of Ordering Nodes}
	%Number of participating ordering nodes in the consensus procedure may also act as an influential factor. Byzantine-fault-tolerant ordering services (e.g., PBFT, BFT-SMaRt), for instance, reach a consensus on a block by exchanging specific messages among the internal peers. This approach may cause a dispute over how many ordering nodes are demanded for the both requirements, that is, lower latency and sufficient security level.
	
	%The mutuality of the whole blockchain parameters is materialized into the performance and latency of a HLF network. Viewed in this light, it may be also possible to present them as mathematical formulas for optimizaiton.

	\subsection{Communicational Adjustments}
	The optimization of communication systems can effectively reduce the AoI. An optimization method worthy of considering may be to design a novel BCE network-specific scheduling policy, different from the existing frameworks. This policy will perform channel allocation to maintain as much latest information as possible in the ledger. %The power control issue in a transmitter level can emerge as a matter to discuss as well. For example, the mimimum power to guarantee that the average AoI does not exceed an acceptable value can be indicated in terms of several practical cases.
	
	In Fig. 5, the AoI sharply increases when the STP is over 0.85. To take precautions against this increase, it is essential to control the high transaction arrival rate of one channel in advance. Inspired by the fact that HLF maintains only one ledger per channel, a multiple ledger policy can be effective to distribute the transaction arrivals over several HLF channels, to eventually reduce the AoI.

	%\subsection{Public Blockchain Cases}
	%A public blockchain network, where any entities can access without approvals, has a different structure from HLF. The main difference is how internal nodes approach consensus. For example, PoW-based public blockchains must contain a miner node set that provides their computational power for data integrity and defect-free blocks, but this is not necessary in a private blockchain. This fact may impede discovering the optimal blockchain parameters. Therefore, a different approach is required for public blockchain cases.

	\section{Conclusions}
	In this article, we explore whether BCE networks can be used for real-time applications. 
	Utilizing blockchain for application data management induces additional latency, which increases the AoI. However, the blockchain is desired for ensuring data integrity without a trusted third party. 
	Therefore, to maintain fresh data in BCE networks, we discuss how to design influential factors on the AoI in BCE networks including the blockchain parameters, the data generation frequency, and the communication parameters. Specifically, the some of the main insights, obtained in this article, can be summarized as follows:
	\begin{itemize}
		\item The optimal values of the block size, the block-generation timeout, and the STP, which minimize the AoI, exist. This is mainly due to the tradeoff that faster block generation (i.e., smaller block size, shorter timeout, and higher STP) reduces the ordering latency, but increases the validation latency.
		\item More frequent generation of the update data is generally expected to be better at retaining fresh data. However, it may not be true in the BCE network because frequent transaction generation results in higher probability of being invalid transactions during the MVCC verification.
	\end{itemize}

	Eventually, this article provides the initial understanding of BEC networks in the data freshness aspects, and paves the way to reliable BEC networks for AoI-sensitive applications and services.


\begin{thebibliography}{99}
	\bibitem{1}
	H.~Lei and D.~Kim, ``Design and implementation of an integrated {I}o{T}
	blockchain platform for sensing data integrity,'' \emph{Sensors}, vol.~19,
	no.~10, pp. 1--26, May 2019.
	
	\bibitem{2}
	Z.~Chen, S.~Chen, H.~Xu, and B.~Hu, ``A security authentication scheme of 5{G}
	ultra-dense network based on block chain,'' \emph{IEEE Access}, vol.~6, pp.
	55\,372--55\,379, May 2018.
	
	\bibitem{3}
	A.~Kosta, N.~Pappas, and V.~Angelakis, ``Age of information: A new concept,
	metric, and tool,'' \emph{Foundations and Trends in Netw.}, vol.~12, no.~3,
	pp. 162--259, Nov. 2017.
	
	\bibitem{4}
	S.~Kaul, R.~Yates, and M.~Gruteser, ``Real-time status: How often should one
	update?'' in \emph{Proc. IEEE Int. Conf. on Comput. Commun. (INFOCOM)},
	Orlando, FL, USA, Mar. 2012, pp. 1--5.
	
	\bibitem{5}
	Y.~Sun, E.~U. Biyikoglu, R.~Yates, C.~E. Koksal, and N.~B. Shroff, ``Update or
	wait: How to keep your data fresh,'' in \emph{Proc. IEEE Int. Conf. on
		Comput. Commun. (INFOCOM)}, San Francisco, CA, USA, Apr. 2016, pp. 1--9.
	
	\bibitem{6}
	E.~Androulaki, A.~Barger, V.~Bortnikov, and C.~Cachin, ``Hyperledger {F}abric:
	A distributed operating system for permissioned blockchains,'' in \emph{Proc.
		Europ. Conf. on Comput. Syst. (EuroSys)}, Porto, Portugal, Jan. 2018, pp.
	1--15.
	
	\bibitem{7}
	J.~Sousa, A.~Bessani, and M.~Vukolic, ``A {B}yzantine fault-tolerant ordering
	service for the {H}yperledger {F}abric blockchain platform,'' in \emph{Proc.
		IEEE/IFIP Int. Conf. on Dependable Syst. and Netw. (DSN)}, Luxembourg City,
	Luxembourg, Jun. 2018, pp. 1--8.
	
	\bibitem{8}
	P.~Thakkar, S.~Nathan, and B.~Viswanathan, ``Performance benchmarking and
	optimizing {H}yperledger {F}abric blockchain platform,'' in \emph{Proc. IEEE
		Int. Symp. on Modeling, Analysis, and Simulation of Comput. and Telecomm.
		Syst. (MASCOTS)}, Milwaukee, WI, USA, Sep. 2018, pp. 1--13.
	
	\bibitem{9}
	M.~B.~H. Weiss, K.~Werbach, D.~C. Sicker, and C.~E.~C. Bastidas, ``On the
	application of blockchains to spectrum management,'' \emph{IEEE Trans. on
		Cognitive Commun. and Netw.}, vol.~5, no.~2, pp. 193--205, Apr. 2019.
	
	\bibitem{10}
	https://github.com/hyperledger/fabric/releases/tag/v1.3.0.
	
	\bibitem{11}
	S.~K. Kaul, R.~D. Yate, and M.~Gruteser, ``Status updates through queues,'' in
	\emph{Proc. Conf. on Inf. Sci. and Syst. (CISS)}, Princeton, NJ, USA, Mar.
	2012, pp. 1--6.
	
	\bibitem{12}
	I.~Kadota, A.~Sinha, and E.~Modiano, ``Optimizing age of information in
	wireless networks with throughput constraints,'' in \emph{Proc. IEEE Int.
		Conf. on Comput. Commun. (INFOCOM)}, Honolulu, HI, USA, Apr. 2018, pp. 1--9.
	
	\bibitem{13}
	Q.~He, D.~Yuan, and A.~Ephremides, ``Optimizing freshness of information: On
	minimum age link scheduling in wireless systems,'' in \emph{Proc. IEEE Int.
		Symp. on Modeling and Optim. in Mobile, Ad Hoc, and Wireless Netw. (WiOpt)},
	Tempe, AZ, USA, May 2016, pp. 1--8.
	
	\bibitem{14}
	Y.~Gu, H.~Chen, Y.~Zhou, Y.~Li, and B.~Vucetic, ``Timely status update in
	internet of things monitoring systems: An age-energy tradeoff,'' \emph{IEEE
		J. Internet of Things}, vol.~6, no.~3, pp. 5324--5335, Jun. 2019.
	
	\bibitem{15}
	J.~Hribar, M.~Costa, N.~Kaminski, and L.~A. Dasilva, ``Using correlated
	information to extend device lifetime,'' \emph{IEEE J. Internet of Things},
	vol.~6, no.~2, pp. 2439--2448, Apr. 2018.
	
	\end{thebibliography}
\end{document}